\RequirePackage{fix-cm}
\documentclass[twocolumn,epjc3]{svjour3}  

\makeatletter\let\cl@chapter\relax\makeatother

\smartqed
\RequirePackage{graphicx}
\usepackage{amsmath,amssymb}
\usepackage{xspace,xcolor,upgreek,enumerate}
\usepackage[amssymb]{SIunits}
\usepackage{cleveref}
\usepackage{comment}

\newcommand{\eerad}{EERAD3\xspace}
\newcommand{\kT}{\ensuremath{k_{\mathrm{T}}}}
\newcommand{\D}{\mathrm{d}\xspace}
\newcommand{\alphaS}{\ensuremath{\alpha_\mathrm{s}}}
\newcommand{\muR}{\ensuremath{\mu_\mathrm{R}}}

\journalname{Eur. Phys. J. C}
\begin{document}

\title{A comparative study of flavour-sensitive observables in hadronic Higgs decays}

\author{Benjamin Campillo Aveleira\thanksref{e1,addr1,addr2}
\and
Aude Gehrmann--De Ridder\thanksref{e2,addr1,addr3}
\and
Christian T Preuss\thanksref{e3,addr1,addr4}
}
\thankstext{e1}{e-mail: benjamin.campillo@kit.edu}
\thankstext{e2}{e-mail: gehra@phys.ethz.ch}
\thankstext{e3}{e-mail: preuss@uni-wuppertal.de}
\institute{Institute for Theoretical Physics, ETH, 8093 Z\"urich, Switzerland \label{addr1}
\and
Institute for Theoretical Physics, KIT, 76131 Karlsruhe, Germany \label{addr2}
\and
Department of Physics, University of Z\"urich, 8057 Z\"urich, Switzerland \label{addr3}
\and
Department of Physics, University of Wuppertal, 42119 Wuppertal, Germany \label{addr4}
}

\date{Received: date / Accepted: date}

\preprint{ZU-TH 12/24, KA-TP-03-2024, P3H-24-011}

\maketitle

\begin{abstract}
Jet production from hadronic Higgs decays at future lepton colliders will have significantly different phenomenological implications than jet production via off-shell photon and $Z$-boson decays, owing to the fact that Higgs bosons decay to both pairs of quarks and gluons.
We compute observables involving flavoured jets in hadronic Higgs decays to three partons at Born level including next-to-leading order corrections in QCD (i.e up to $\mathcal{O}(\alphaS^2))$. The calculation is performed in the framework of an effective theory in which the Higgs boson couples directly to gluons and massless $b$-quarks retaining a non-vanishing Yukawa coupling. 
For the following flavour sensitive observables: the energy of the leading and subleading flavoured jet, the angular separation and the invariant mass of the leading $b$-$\bar{b}$ pair, we contrast the results obtained in both Higgs decay categories and using either of the infrared-safe flavoured jet algorithms flavour-$\kT$ and flavour-dressing. 
\end{abstract}

\section{Introduction}
\label{sec:introduction}
Precision studies of the Higgs boson discovered at LHC by CMS and ATLAS \cite{ATLAS:2012yve,CMS:2012qbp} will become possible at future lepton colliders such as \cite{Abada:2019zxq,Moortgat-Picka:2015yla}, all aiming to operate as Higgs factories. In this clean experimental environment, where interactions take place at well-defined centre-of mass energies, it is expected to enable model-independent measurements of the Higgs couplings to gauge bosons and fermions at the level of a few percent. At these future lepton colliders, in particular it will become possible to have access to so-far unobserved hadronic decay channels such as Higgs decays to gluons. The latter is currently 
inaccessible in hadron-collider environments due to the presence of overwhelming QCD backgrounds. Only the $H\to b\bar{b}$ decay was observed to date \cite{Aaboud:2018zhk,Sirunyan:2018kst} in associated vector-boson production where the leptonic decay signature of the vector boson helps to identify the $H \to b\bar{b}$ decay.

Hadronic Higgs decays to at least two final state hard partons proceed via two main decay modes; either as Yukawa-induced decay to a bottom-quark pair, $H\to b \bar b$, or as a heavy-quark-loop induced decay to two gluons, $H\to gg$. In the latter category, observables are computed in the framework of an effective field theory, in which the top-quark loop is integrated out into an effective point-like $Hgg$ vertex.

So far these two categories of Higgs decay processes have been considered together in the computation of flavour-agnostic event-shape observables, i.e., for three-jet-like final states  in \cite{Coloretti:2022jcl,Gao:2016jcm,Gao:2019mlt,Luo:2019nig,Gao:2020vyx} and for four jet-like
final states in \cite{Gehrmann-DeRidder:2023uld}. It was also recently suggested to determine branching ratios in hadronic Higgs decays via fractional energy correlators \cite{Knobbe:2023njd}.
Flavour-sensitive jet observables related to the presence of a flavoured jet in the final state have so far been computed for the following LHC processes: VH production, with $H\to b\bar b$  or $Z+b$-jet and $Z/W+c$-jet, with the vector boson decaying leptonicallly in all cases.
More precisely, parton-level predictions including up to NNLO QCD corrections using massless charm or bottom quarks at the origin of the flavoured jet have been computed most recently for $VH$ in \cite{Gauld:2019yng,Bizon:2019tfo,Zanoli:2021iyp,Behring:2020uzq}, and for $Z/W+c/b$ in \cite{Gauld:2020deh,Czakon:2020coa,Czakon:2022khx,Gauld:2023zlv,Gehrmann-DeRidder:2023gdl,Fedkevych:2022mid}. 

In a lepton collider environment, flavour-sensitive jet observables associated to $Z$ decays have been lately considered in \cite{Gauld:2022lem,Caola:2023wpj}
while colour-sensitive observables have been used to disentangle bottom quarks stemming from Higgs decays against those stemming fror QCD background in \cite{Cavallini:2021vot}.

All of these computations employ an infrared-safe procedure to define flavoured jets from massless partons.  
The latter procedure requires the use of an infrared-safe 
recombination algorithm to cluster flavourless and flavoured partons into final states including well-defined flavoured jets. 
Up to very recently, only the flavour-$\kT$ algorithm \cite{Banfi:2006hf} was used in theoretical computations at hadron colliders.
Lately, a number of flavour-sensitive jet algorithms have been designed \cite{Gauld:2022lem,Caola:2023wpj,Caletti:2022hnc,Czakon:2022wam}, with increased interest in providing modified versions of the anti-$\kT$ algorithm  \cite{Gauld:2022lem,Caola:2023wpj,Czakon:2022wam}, to improve the data versus theory comparisons at LHC. 
%%
\begin{comment}
%In contrast, dedicated implementations needed for two- and %three-jet production associated to off-shell photon or $Z$-%boson decays have explicitly only been considered using the %flavour-$\kT$ algorithm \cite{Banfi:2006hf} and the flavour-%dressing algorithm \cite{Gauld:2022lem}. 
\end{comment}

All of these algorithms share the principle that, at least up to a certain order in the strong coupling, they can be proven to be infrared flavour safe, i.e., that the flavour assignment of a given jet is not spoilt by the emission of unresolved (soft or collinear) massless partons. A detailed comparison of currently available algorithms has been presented in \cite{Caola:2023wpj}.

In this paper, we compute a range of flavour-sensitive three-jet like observables in hadronic Higgs decays, including NLO QCD corrections for both decay categories, i.e related to $H\to b \bar b$ and $H\to gg$, as alluded to above. 
To define flavour-sensitive observables we employ the flavour-$\kT$ algorithm \cite{Banfi:2006hf} and the flavour-dress-ing algorithm \cite{Gauld:2022lem} which can both be applied in a lepton collider environment. 
 
The study is organised as follows: 
In \cref{sec:flavoured_jets} we give a brief overview of the definition of flavoured jets using either of the two infrared flavour-safe jet algorithms
before describing our computational setup in \cref{sec:computational_setup}. In \cref{sec:results}, in a first part presented in \cref{subsec:flavour_safety} we compute the so-called ``misindentified cross section'' to check the correctness of our implementation using both infrared-safe and flavour-safe jet algorithms. 
In the second part of \cref{sec:results}, i.e., in \cref{subsec:predictions}, we present theoretical predictions up to second order in QCD for four different flavour-sensitive observables related to both Higgs-decay categories and compare the results using both flavour-safe jet algorithms. We conclude and give an outlook on future work in \cref{sec:conclusions}.

\section{Flavoured jets}\label{sec:flavoured_jets}
In experimental analyses as well as theoretical calculations, final-state particle configurations are often described by so-called jets. While the definition of jets is fixed by the choice of a sequential recombination algorithm known as a jet algorithm, the association of the jet with a given parton (or hadron) flavour is a more complicated endeavour.
Naïvely, it may be tempting to define the jet flavour as the sum of the constituent flavours of each jet, i.e,
\begin{equation}
    \text{Jet flavour} = {f}-{\bar{f}} \, ,
\label{eq:naïve_jetflav}
\end{equation}
where $f$ is the number of particles with a given a flavour and $\bar{f}$ is the number of particles with the corresponding anti-flavour.
However, this naïve approach violates an important property known as infrared flavour safety \cite{Banfi:2006hf}, which describes whether an algorithm respects physically meaningful flavour assignments when one or more particles become unresolved.
The problem is that jet algorithms may in general cluster the flavoured daughters of a soft wide-angle gluon splitting $g \to q\bar q$ into different jets, changing their flavour association.
In effect, the flavours of the two jets, and therefore any flavour-dependent observables, become explicitly dependent on the presence of a pair of unresolved particles, in violation of infrared safety.
It is thus mandatory for a jet-flavour definition to respect infrared flavour safety, meaning that the presence of a pair of unresolved fla-voured particles to an event must not change the flavours of the jets.

As mentioned in the introduction, 
several approaches to infrared-safe flavoured jet algorithms have been explored \cite{Gauld:2022lem,Caola:2023wpj,Banfi:2006hf,Caletti:2022hnc,Czakon:2022wam}, of which only the flavour-$\kT$ \cite{Banfi:2006hf} and the flavour-dressing algorithm \cite{Gauld:2022lem} explicitly describe an implementation compatible with the use of the Durham ($\kT$) algorithm. In the present study we therefore limit the discussion to these two flavoured jet algorithms, which we briefly describe below.

The flavour-$\kT$ algorithm \cite{Banfi:2006hf} achieves infrared fla-vour safety by modifying the Durham distance measure valid for all unflavoured partons $(i,j)$ \cite{Catani:1991hj} into two cases defined depending whether the softer particles to be clustered are flavoured or not. 
In the flavour-$\kT$ algorithm, the distance measure is given by: 
\begin{equation}
    y_{ij}^{\mathrm{F}} = \frac{2(1-\cos\theta_{ij})}{E_{\mathrm{tot}}^2} \min(E_i,E_j)^{2-\alpha}\max(E_i,E_j)^{\alpha}
\label{eq:distmeas_flavour}
\end{equation}
if the softer of $i$, $j$ is flavoured, Instead the regular Durham distance given by,
\begin{equation}
    y_{ij}^{\mathrm{F}} = \frac{2(1-\cos\theta_{ij})}{E_{\mathrm{tot}}^2} \min(E_i,E_j)^2 \, ,
\end{equation}
is kept if the softer of $i$, $j$ is flavourless. 
Here, $\alpha \in (0,2]$ is an arbitrary parameter.
In this study, the value $\alpha=2$ is chosen everywhere.
The distance measure in \cref{eq:distmeas_flavour} ensures that soft pairs of flavoured particles are recombined first and thus avoids the previously discussed infrared-safety problem present in the standard (flavour-agnostic) Durham ($k_T$) algorithm.
The main draw-back to the use of this algorithm is that it requires the flavour information of all particles, thus making it difficult to use in experimental analyses. Indeed, so far, it has not be used  
in measurements of flavour-sensitive jet observables.

The flavour-dressing algorithm \cite{Gauld:2022lem} on the other hand does not change the underlying jet algorithm but instead alters the way flavours are assigned to jets. Instead of using the naïve jet-flavour definition, a more complex definition is invoked which ensures infrared flavour safety of the jet algorithm at hand. The basic idea is to first cluster an event into jets using a flavour-agnostic jet algorithm of choice, and, in a second step, cluster flavoured objects into so-called flavour clusters using a technique akin to soft-drop grooming \cite{Larkoski:2014wba}. In the final, ``association'' step the flavour clusters of step two are assigned to the jets of step one. A striking and important feature of the flavour-dressing algorithm is that it is equally applicable to theoretical calculations (and full particle-level simulations) as to experimental analyses, as discussed in \cite{Gauld:2022lem}.
It can in particular be used together with the anti-$k_T$ algorithm, mostly used in experimental analyses at LHC
but also in combination with the Durham $k_T$ algorithm 
as explored in this paper.

\begin{figure*}
    \centering
    \includegraphics[width=0.46\textwidth]{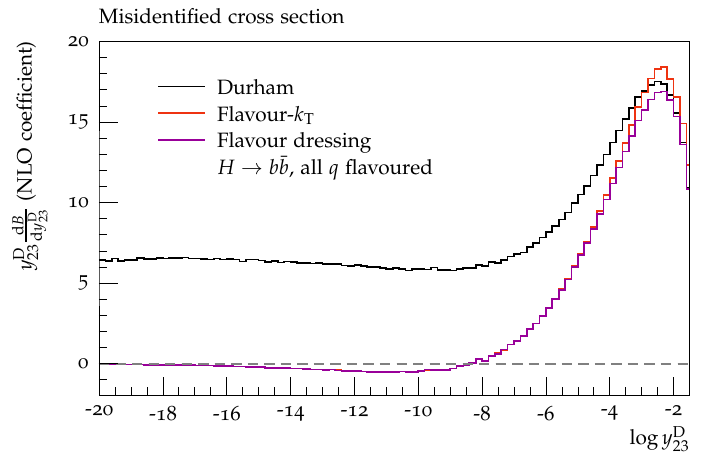}
    \includegraphics[width=0.46\textwidth]{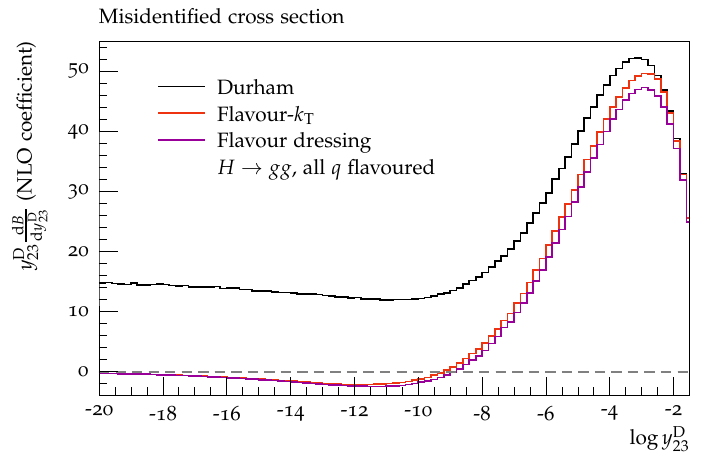}\\
    \includegraphics[width=0.46\textwidth]{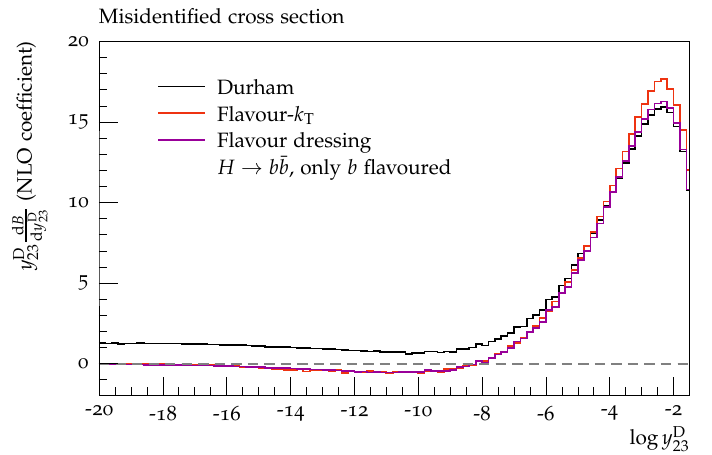}
    \includegraphics[width=0.46\textwidth]{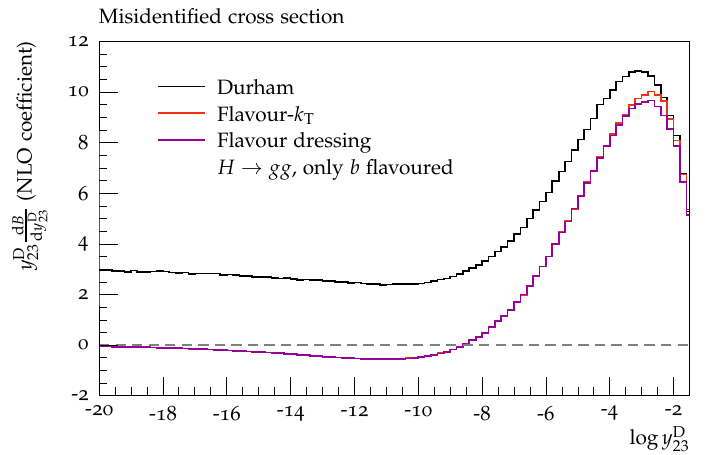}
    \caption{Misidentified cross section in the $H\to b\bar{b}$ (left) and $H\to gg$ (right) category, considering all quarks as flavoured (top row) and only $b$-quarks as flavoured (bottom row). Both flavour-$\kT$ and flavour dressing use $\alpha = 2$, as defined in the text.}
    \label{fig:misid_xsec}
\end{figure*}

\section{Computational setup}\label{sec:computational_setup}
We perform the computation of flavoured observables related to hadronic Higgs decays using the parton-level Monte-Carlo generator \eerad \cite{Gehrmann-DeRidder:2007vsv,Gehrmann-DeRidder:2014hxk}, which was originally developed to compute NNLO QCD corrections to event-shape observables in hadronic $Z$-decays.
This generator was recently extended to compute event sha-pes related to hadronic Higgs decays with three- and four-jet configurations at Born level in \cite{Coloretti:2022jcl,Gehrmann-DeRidder:2023uld}. In both cases, the antenna subtraction method is used to regulate infrared divergences related to real radiation contributions.

The hadronic Higgs-decay observables are computed in the framework of an effective theory including two Higgs decay categories. In the first category, the Higgs boson couples directly to gluons via an effective Higgs-gluon-gluon vertex and in the other, Standard-Model-like category, the Higgs boson decays into a massless $b$-quark pair retaining a non-vanishing Yukawa coupling \cite{Coloretti:2022jcl,Gehrmann-DeRidder:2023uld}. This can be recasted into the following effective Lagrangian
\begin{equation}
    \mathcal{L}_\mathrm{Higgs} = -\frac{\lambda(M_t,\muR)}{4}HG_{\mu\nu}^aG^{a,\mu\nu} + \frac{y_b(\muR)}{\sqrt{2}}H\bar{\psi}_b\psi_b \, .
\label{eq:lagrangian}
\end{equation}
In this context, the effective Higgs-gluon-gluon coupling is given in terms of the Higgs vacuum expectation value $v$ by
\begin{equation}
    \lambda(M_t,\muR) = -\frac{\alphaS(\muR)C(M_t,\muR)}{3\uppi v}
\end{equation}
and the $Hb\bar{b}$ Yukawa coupling reads
\begin{equation}
    y_b(\muR) = \bar{m}_b(\muR)\frac{4\uppi\alpha}{\sqrt{2}M_W\sin\theta_\mathrm{W}} \, .
\end{equation}
Both $\lambda$ and $y_b$ are subject to renormalisation, which we perform at scale $\muR$ in the $\overline{\text{MS}}$ scheme using $N_\mathrm{F} = 5$. The top-quark Wilson coefficient is evaluated at first order in $\alphaS$ using the results of \cite{Inami:1982xt,Djouadi:1991tk,Chetyrkin:1997iv,Chetyrkin:1997un,Chetyrkin:2005ia,Schroder:2005hy,Baikov:2016tgj}, and the running of the Yukawa mass $\bar{m}_b$ is performed using the results of \cite{Vermaseren:1997fq}.

It is important to highlight that the terms in \cref{eq:lagrangian} do not interfere under the assumption of kinematically massless quarks. In particular, they do not mix under renormalisation \cite{Gao:2019mlt}. This allows to define two separate Higgs-decay categories and to compute higher-order corrections independently for each. Throughout, we therefore consider predictions for the $H\to b\bar{b}$ and $H\to gg$ categories separately.
All partonic contributions yielding three hard partons in the final state at Born level and needed for the computation presented in this paper, have been presented in detail for both categories up to $\mathcal{O}(\alphaS^2)$ in \cite{Coloretti:2022jcl}. 
In particular, it is worth mentioning that the Born-level partonic processes contributing at $\mathcal{O}(\alphaS)$ are: $H\to b \bar b g$ in the $H\to b \bar b$ category and $H\to ggg$, $H\to g q \bar{q}$ in the $H\to gg$ category. In the latter case, $q$ stands for any quark with specific flavour, including the bottom quark. 

For any infrared-safe observables $O$, the parton-level generator \eerad calculates the LO coefficient $\bar A$ and the NLO coefficient $\bar B$ in a perturbative expansion of the differential decay width,
\begin{multline}
    \frac{1}{\Gamma^{(n)}_{2j}(\muR)}\frac{\mathrm{d}\Gamma}{\mathrm{d} O} = \left(\frac{\alphaS(\muR^2)}{2\uppi}\right)\frac{\D \bar A}{\D O} \label{eq:decayWidth} \\
    + \left(\frac{\alphaS(\muR^2)}{2\uppi}\right)^2\left(\frac{\D \bar B}{\D O} + \beta_0\log\left(\frac{\muR^2}{M^2}\right)\frac{\D \bar A}{\D O}\right) \, .
\end{multline}
Here, $\Gamma^{(n)}_{2j}$ is the partial two-body decay width to order $n$; specifically, $n=0$ at LO and $n=1$ at NLO.

In \cref{eq:decayWidth}, the LO coefficient $\bar A$ involves only an integration over the three-particle phase space, while the NLO contribution $\bar B$ involves an integration over the four-particle phase space related to the real-radiation contribution and real subtraction terms, and an integration over the three-particle phase space, pertaining to one-loop contributions and virtual subtraction terms. 
Within this study, we employ the antenna subtraction scheme to construct real and virtual subtraction terms.
Explicit expressions for the perturbative coefficients $\bar A$ and $\bar B$ can be found in \cite{Coloretti:2022jcl}.

In order to compute 
flavour-sensitive observables related to hadronic Higgs decays, as in this paper, on top of the ingredients needed to compute flavour-agnostic observables as described above, a new flavour layer needs to be implemented in  the parton-level event generator \eerad.
This parton-level flavour-tracking procedure implemented here in \eerad for the first time can be summarized as follows:
For each momentum configuration all contributing flavour configurations are generated and matrix elements as well as subtraction terms are evaluated separately for each flavoured parton configuration.
In particular, this means that the calculation is split into different flavour contributions across all layers in \cref{eq:decayWidth}.
Because the subtraction terms involve mapped configurations in which one parton is clustered into a ``reduced'' particle configuration, this has the consequence that the subtraction scheme is only viable if a flavour-safe jet algorithm is used.
For each flavour configuration, the flavour layer then acts as an additional input to the jet algorithm and parton-level contributions needed to be considered for the evaluation of the observables. 

To conclude this section, we wish to define the theoretical framework in which our predictions are valid and numerically stable.
Fixed-order calculations are accurate only in phase-space regions in which hard, well-separated jets dominate.
We therefore devise a resolution cut,  $y_\mathrm{cut}$ on the (flavoured) jet algorithm and define three-jet states to have three particles with $\min_{i,j}(y_{ij}) \geq y_\mathrm{cut}$, where $y_{ij}$ denotes the distance measure of the respective jet algorithm. Similarly, four-jet states are defined to have four particles with $\min_{i,j}(y_{ij}) \geq y_\mathrm{cut}$.
Furthermore, to avoid large numerical cancellations between the real contribution and the corresponding real subtraction term in unresolved phase-space regions, we implement a technical cut-off of $y_0 = 10^{-8}$ on the smallest dimensionless two-particle invariant $y_{ij} = 2p_ip_j/s$ in real configurations. We have verified that our predictions are independent of the choice of this theoretical cut-off.

\begin{figure}
    \centering
    \includegraphics{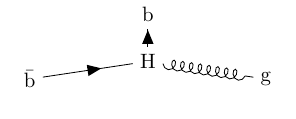}
    \caption{Momentum-space diagram in the Higgs rest frame of a $H\to b\bar b g$ decay in which flavour-$\kT$ allows for an arbitrarily soft $b$-quark.}
    \label{fig:Hbbg}
\end{figure}

\section{Results}\label{sec:results}
We here focus on presenting theoretical predictions for the following flavour-sensitive observables: the energy of the leading and subleading flavoured jet, the angular separation and invariant mass of the leading $b$-$\bar{b}$ pair. We shall present results obtained in both Higgs-decay categories and using either of the infrared-safe flavoured jet algorithms flavour-$\kT$ and flavour-dressing. 
After presenting the numerical set-up in \cref{subsec:Numerical set-up} we divide the discussion of our results into two subsections. In \cref{subsec:flavour_safety} we validate the infrared flavour safety of the flavour-$\kT$ and flavour-dressing algorithms before presenting and comparing our predictions for flavour-sensitive observables in \cref{subsec:predictions}. 

\subsection{Numerical set-up and scale-variation prescription}
\label{subsec:Numerical set-up}
We consider on-shell Higgs decays with $M_H = 125~\giga e\volt$ 
%with a centre-of-mass energy of $M_H = 125~\giga e\volt$ 
and calculate all observables in the resonance centre-of-mass frame. 
We use $\alphaS(M_Z) = 0.1179$ with either one- or two-loop running at LO and NLO, respectively.
Electroweak quantities are considered constant and we use the $G_\mu$-scheme with 
\begin{align}
    G_\mathrm{F} &= 1.20495\times 10^{-5}~\giga e\volt^{-2} \, , \nonumber \\
    M_Z &= 91.1876~\giga e \volt \, , \\
    M_W &= 80.385~\giga e\volt \, , \nonumber
\end{align}
corresponding to $\alpha=\frac{1}{128}$.
To estimate theoretical uncertainties from missing higher-order corrections in our calculation, we vary the renormalisation scale around the Higgs mass, i.e., consider $\muR = k_\mu M_H$ with $0.5 \le k_\mu \le 2$. 

%% Discussion of misid
\subsection{Infrared flavour safety}
\label{subsec:flavour_safety}
To guarantee infrared flavour safety for flavour-sensitive observables, the jet algorithms that are employed in the computation need to correctly assign flavours to jets in the deep infrared region. Correct flavour assignment is determined by the underlying two parton process, i.e., by either two flavourless jets in the $H \to gg$ category or by one flavoured and one anti-flavoured jet in the $H \to b \bar b$ category. 

The study of infrared-flavour safety was first condu-cted for hadronic $Z$-decays in \cite{Banfi:2006hf}, in which it was highlighted that the Durham algorithm violates infrared flavour safety and a flavour-safe modification in terms of the flavour-$\kT$ algorithm was suggested. As a measure of infrared flavour safety, the so-called misidentified cross section was defined in terms of the three-jet resolution variable $y_{23}$. The latter variable, measures the departure from two-jet-like into three-jet-like topologies.

In \cite{Gauld:2022lem} the same criterion was used to validate the infrared flavour safety of the flavour-dressing algorithm. 
As a validation of our implementation, we thus apply the same criterion but here for hadronic Higgs decays. 
In this case, the misidentified cross section collects two-jet like events for which the flavour does not correspond to the flavour of the original two parton 
like topology in either of the Higgs categories. 
In \cref{fig:misid_xsec}, we present the NLO coefficient $\bar B$ differential in the Durham resolution $y_{23}^\mathrm{D}$ using jet definitions according to the plain Durham, flavour-$\kT$, and flavour-dressing algorithm.
We employ two different flavour definitions; in the top row all quarks are counted as flavoured, in line with the definitions used in \cite{Gauld:2022lem,Banfi:2006hf}, while in the bottom row only $b$-quarks are treated as flavoured with all other quarks being flavourless. The former (in $H\to b \bar b$) can be used as a fundamental cross-check of our implementation with the original papers \cite{Gauld:2022lem,Banfi:2006hf}, whereas the latter is the flavour definition used in the remainder of this paper, since it allows for a straightforward experimental adaption in terms of $b$-tagging, see e.g. \cite{CMS:2017wtu,ATLAS:2015thz,LHCb:2015tna}.
In \cref{fig:misid_xsec},  to probe the deep infrared region, we have used the value of the theoretical cut $y_{0}$ 
to be $10^{-13}$.  

For an infrared-safe flavour jet algorithm, one ex-pects the probability for flavour misidentification to vanish as the variable $y_{23}$ tends to zero.
In both Higgs-decay categories and regardless of the flavour-definition, in \cref{fig:misid_xsec} it is clearly visible that the Durham algorithm has a non-zero probability for an incorrect flavour assignment in the deep infrared region, whereas both the flavour-$\kT$ and flavour-dressing algorithm yield a vanishing cross section for misidentified events. Confirming the findings in \cite{Gauld:2022lem,Banfi:2006hf}, both the flavour-$\kT$ and flavour-dressing algorithms provide infrared-safe flavour jet definitions in hadronic Higgs decays, whereas the flavour-agnostic Durham algorithm does not.
Thus, we shall present predictions only using these two jet algorithms in the remainder of this section.

\begin{figure*}
    \centering
    \includegraphics[width=0.48\textwidth]{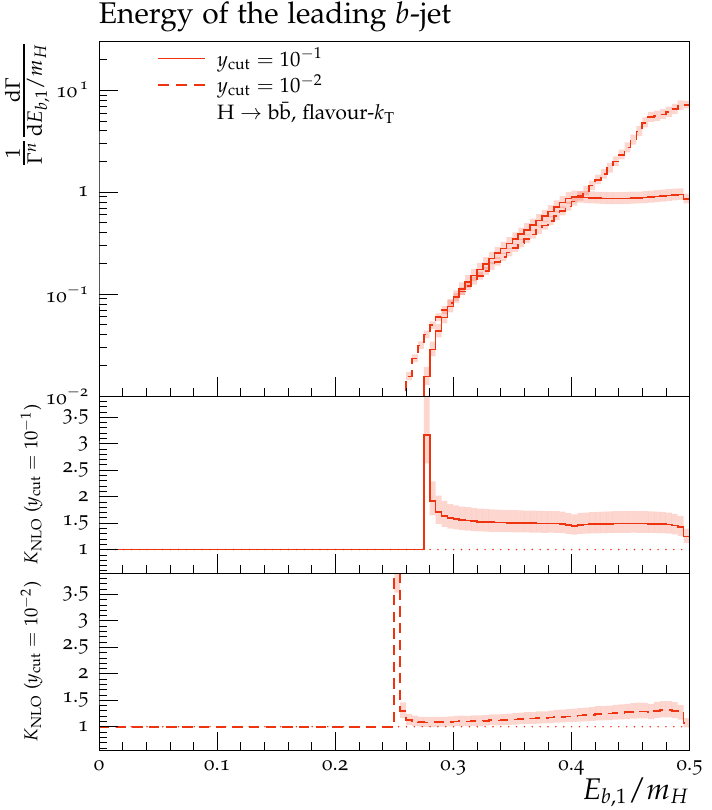}
    \includegraphics[width=0.48\textwidth]{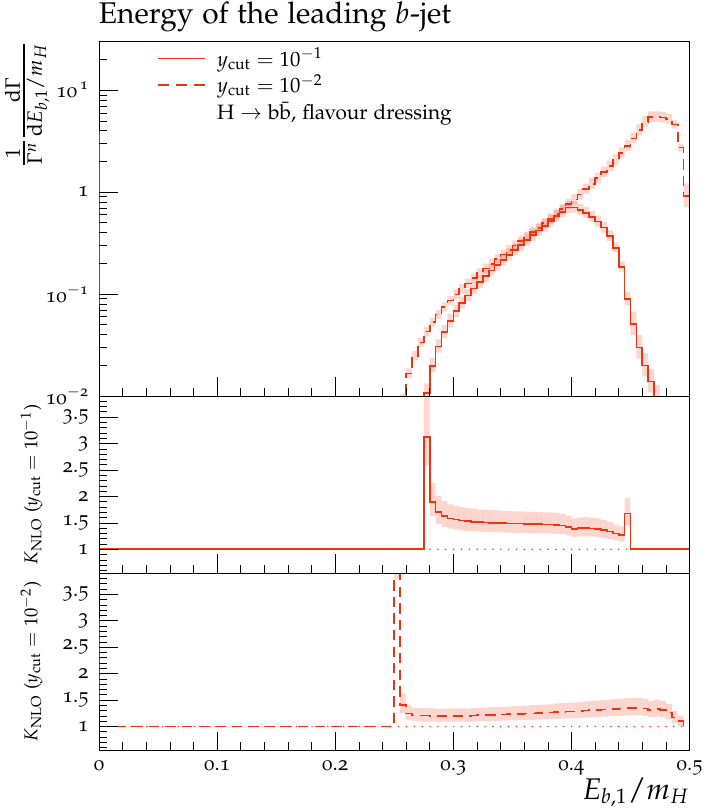}\\
    \includegraphics[width=0.48\textwidth]{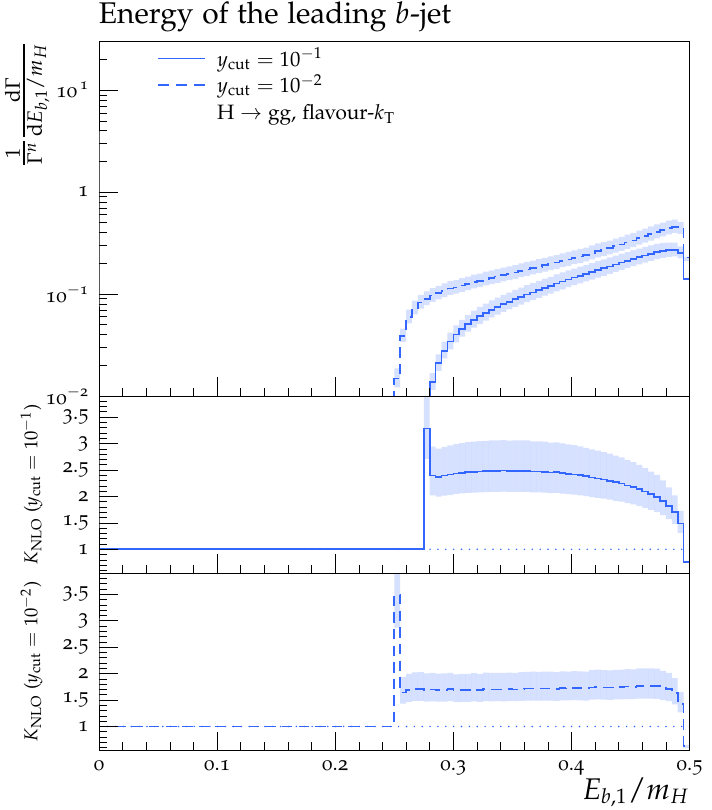}
    \includegraphics[width=0.48\textwidth]{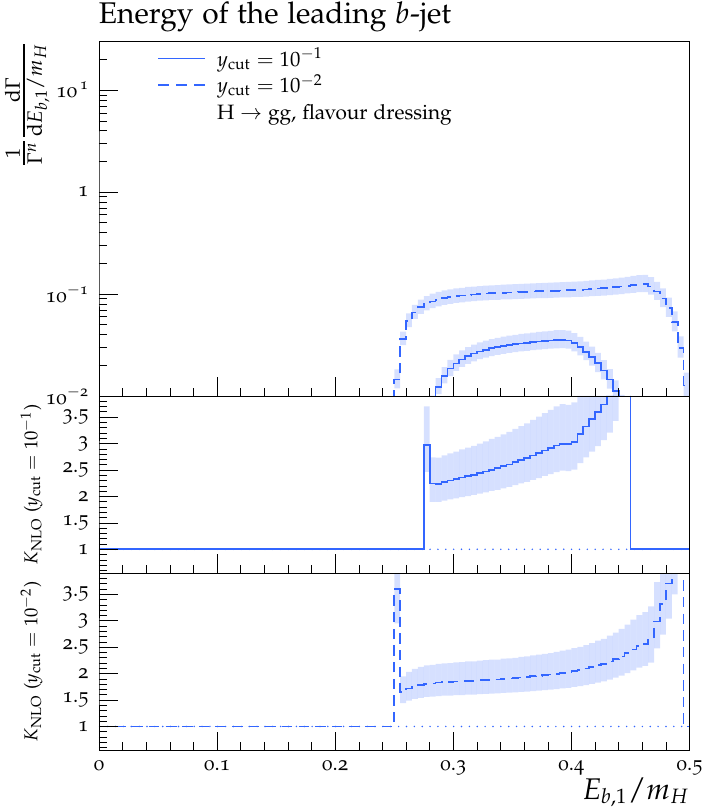}
    \caption{Energy of the leading $b$-jet in the $H\to b\bar{b}$ (\textit{top row}) and $H\to gg$ (\textit{bottom row}) decay category using the flavour-$\kT$ (\textit{left column}) and flavour dressing (\textit{right column}) algorithm.}
    \label{fig:Ef1}
\end{figure*}

\subsection{Flavour-sensitive observables}
\label{subsec:predictions}
%% to be more specific about the three jet configurations here 
We consider flavour-sensitive observables in both Higgs decay categories yielding three-jet-like configurations at Born level. For a given experimental resolution parameter  
$y_\mathrm{cut}$ and  
for each pair of final state partons $i,j$ (quark or gluon), present in the partonic subprocesses at leading and next-to-leading order, we impose that 
$\min_{i,j}(y_{ij}) \geq y_\mathrm{cut}$
in the respective flavoured jet algorithm. We then calculate NLO QCD, i.e ($\mathcal{O}(\alphaS^2)$), predictions for the following observables
\begin{enumerate}[(a)]
\item the energy of the leading flavoured jet, $E_{b,1}$;
\item the energy of the subleading flavoured jet, $E_{b,2}$;
\item the angular separation of the leading $b$- and $\bar b$-jet, $\cos\theta_{b\bar b}$;
\item the invariant mass of the leading $b$- and $\bar b$-jet, $m_{b\bar b}$.
\end{enumerate}
Except for the angular separation, we only consider scaled observables, normalised to the Higgs mass $m_H$. In all cases, we choose $\alpha = 2$ in the flavour-$\kT$ and flavour-dressing algorithm and consider only $b$-quarks 
($\bar{b}$-quarks) as flavoured (anti-flavoured).

\begin{figure*}
    \centering
    \includegraphics[width=0.48\textwidth]{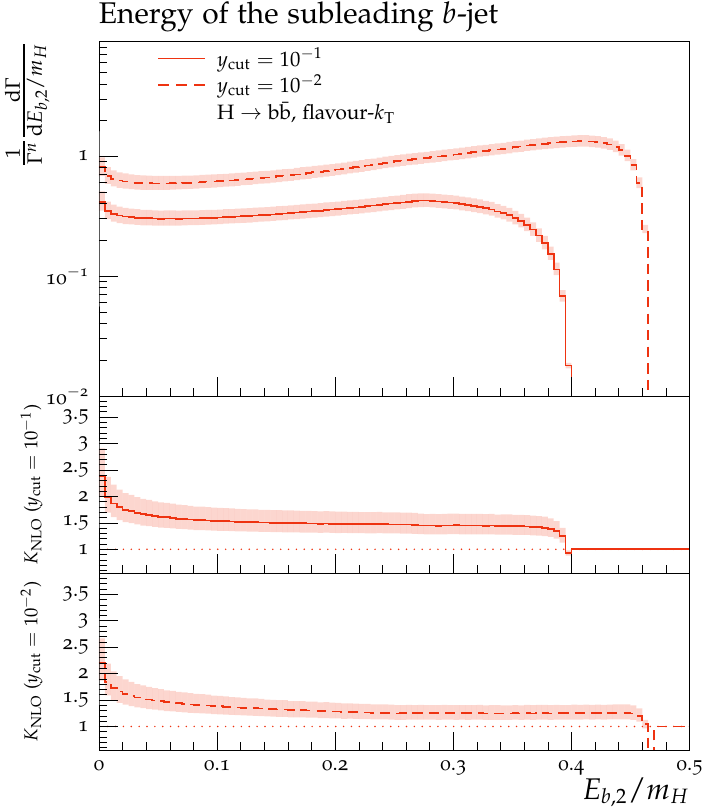}
    \includegraphics[width=0.48\textwidth]{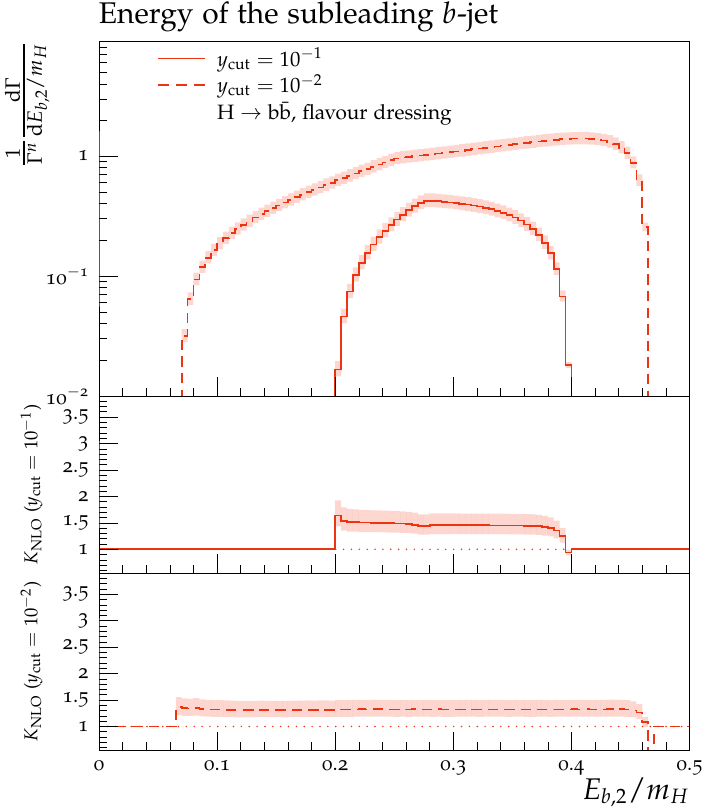}
    \includegraphics[width=0.48\textwidth]{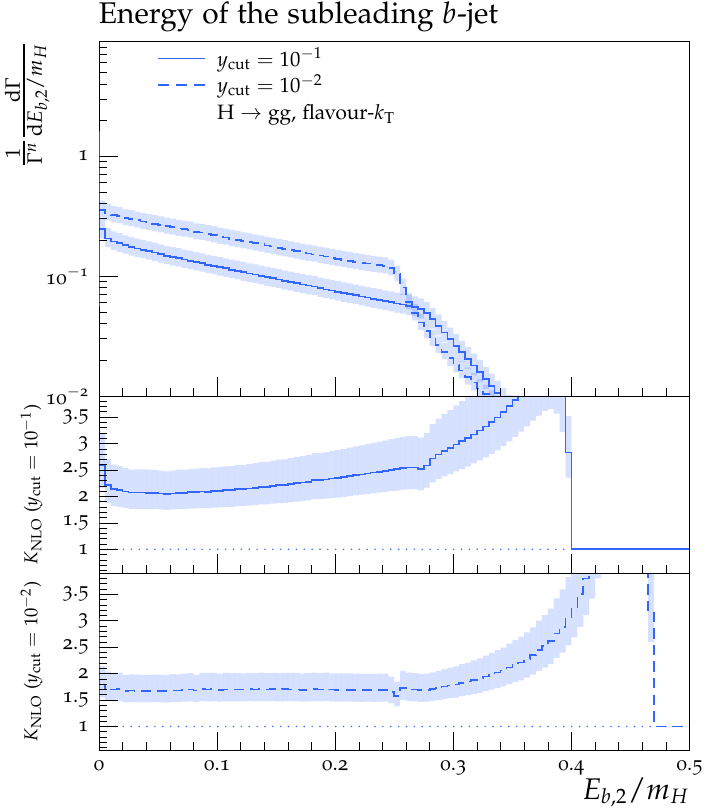}
    \includegraphics[width=0.48\textwidth]{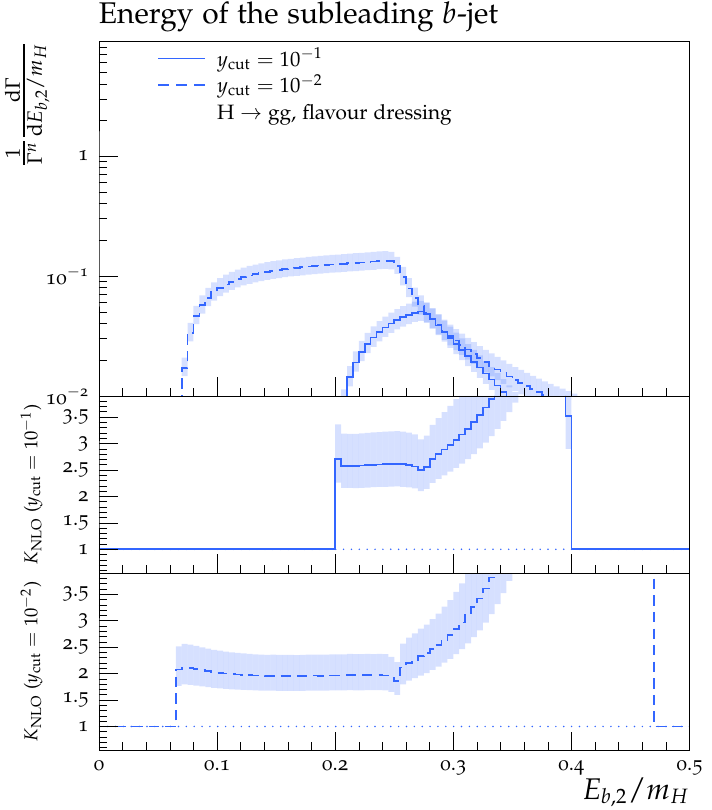}
    \caption{Energy of the subleading $b$-jet in the $H\to b\bar{b}$ (\textit{top row}) and $H\to gg$ (\textit{bottom row}) decay category using the flavour-$\kT$ (\textit{left column}) and flavour dressing (\textit{right column}) algorithm.}
    \label{fig:Ef2}
\end{figure*}

\begin{figure*}
    \centering
    \includegraphics[width=0.48\textwidth]{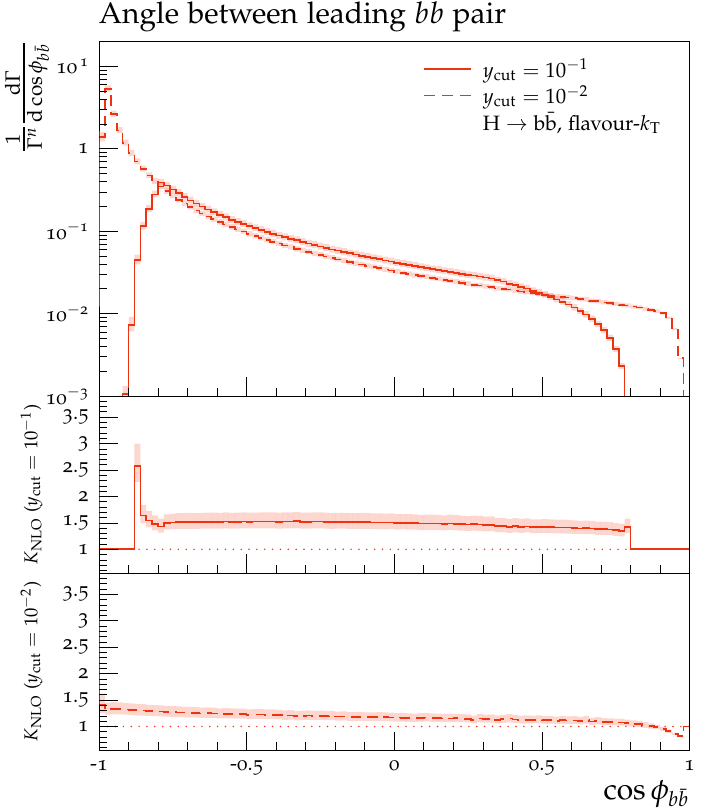}
    \includegraphics[width=0.48\textwidth]{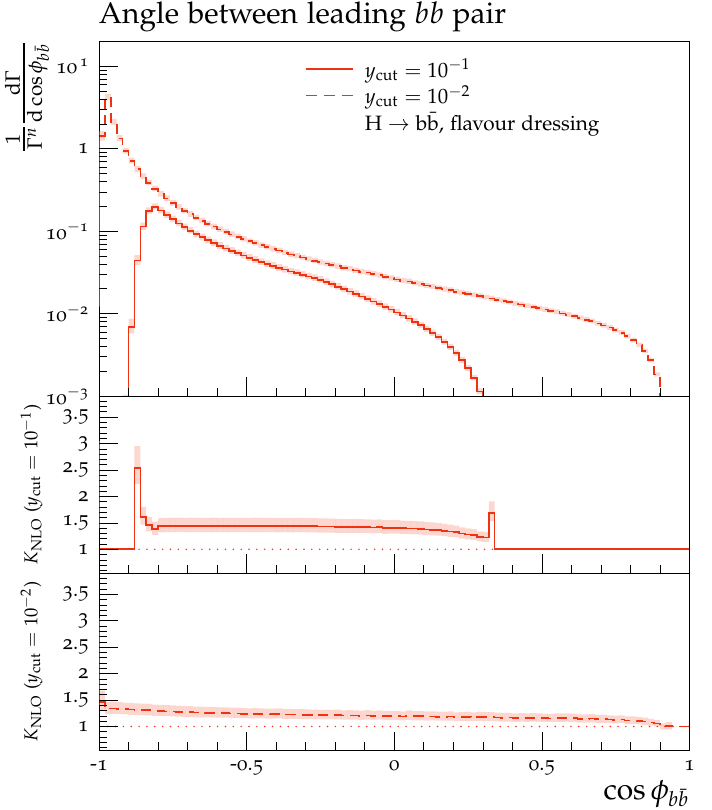}
    \includegraphics[width=0.48\textwidth]{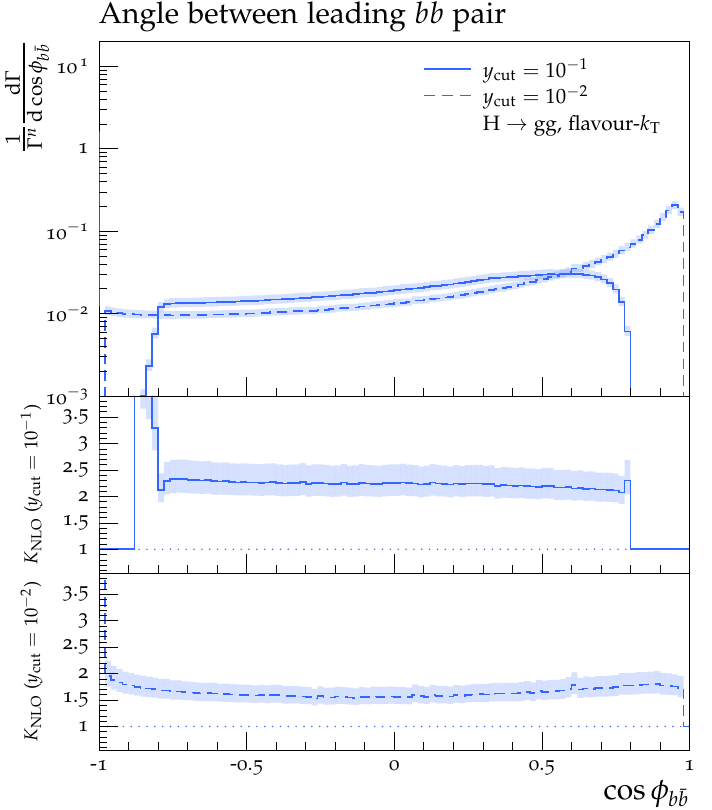}
    \includegraphics[width=0.48\textwidth]{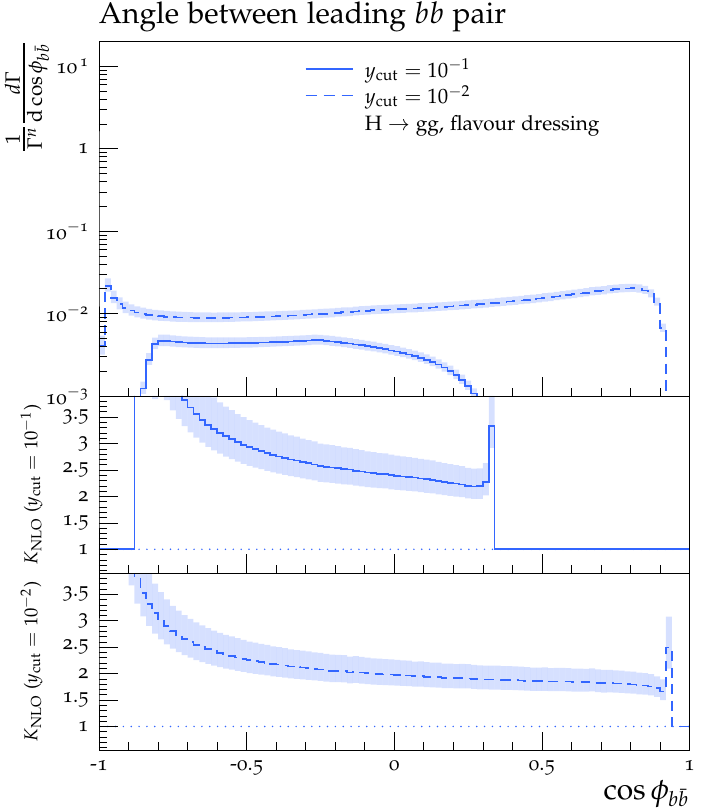}
    \caption{Angular separation of the leading $b$- and $\bar{b}$-jet pair in the $H\to b\bar{b}$ (\textit{top row}) and $H\to gg$ (\textit{bottom row}) decay category using the flavour-$\kT$ (\textit{left column}) and flavour dressing (\textit{right column}) algorithm.}
    \label{fig:cosphi}
\end{figure*}

\begin{figure*}
    \centering
    \includegraphics[width=0.48\textwidth]{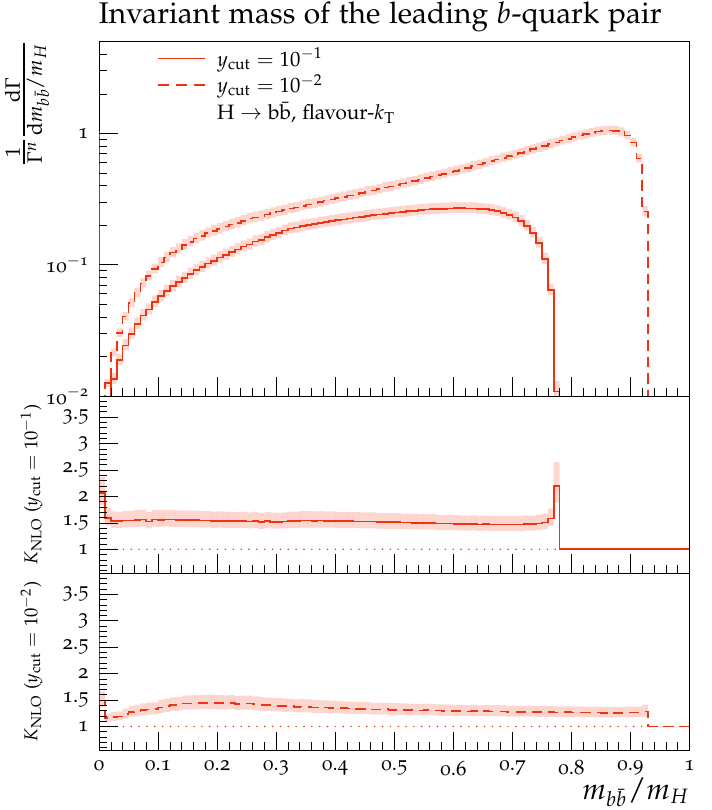}
    \includegraphics[width=0.48\textwidth]{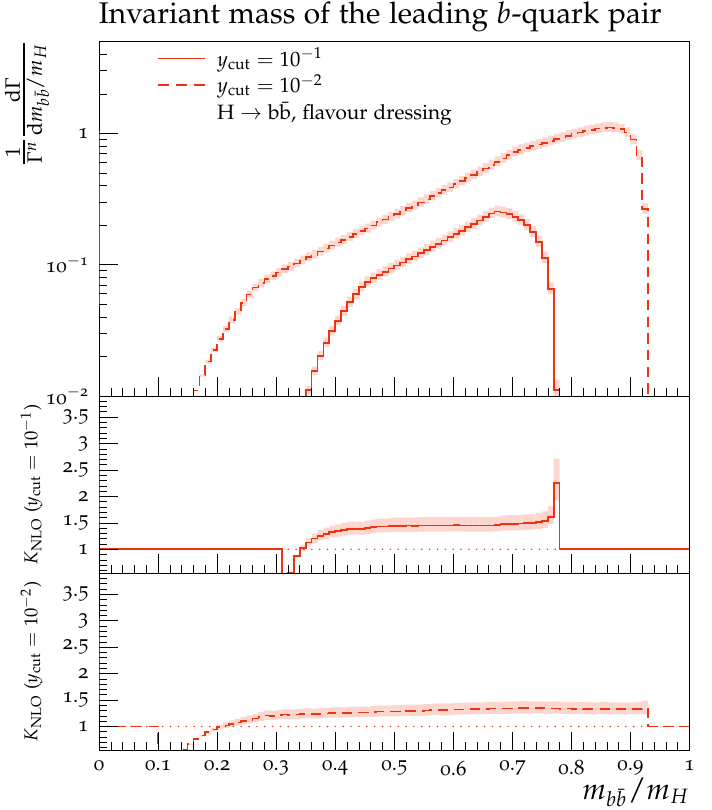}
    \includegraphics[width=0.48\textwidth]{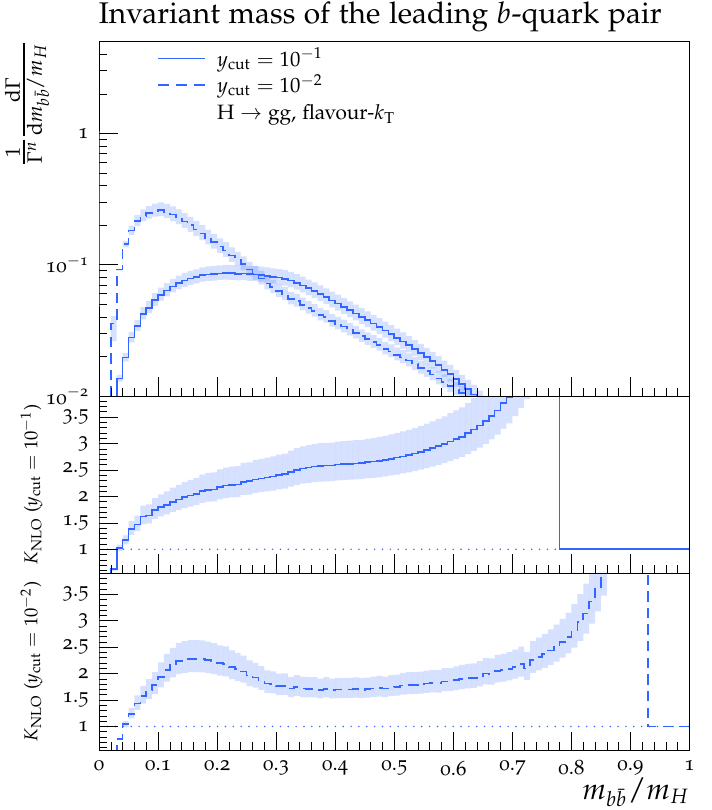}
    \includegraphics[width=0.48\textwidth]{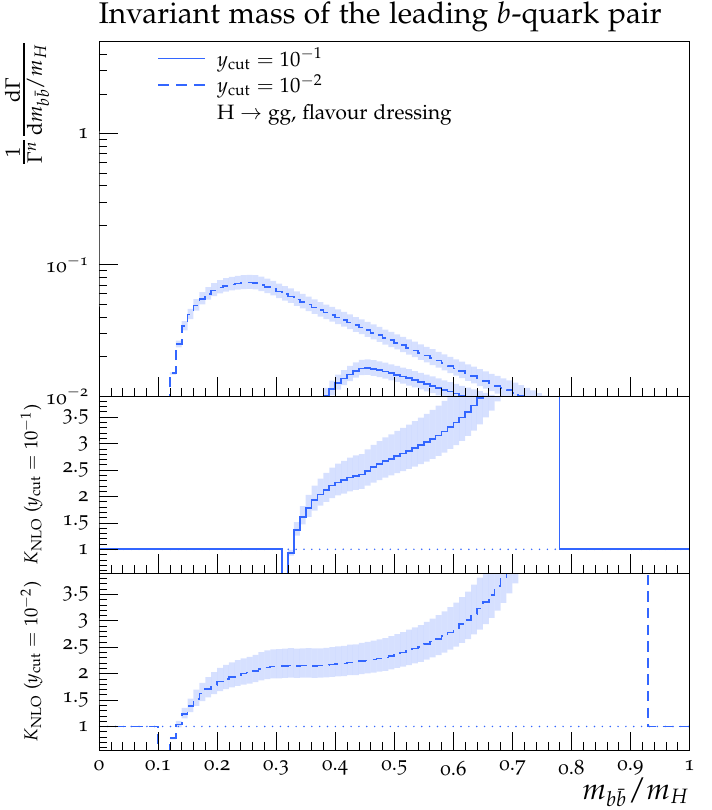}
    \caption{Invariant mass of the leading $b$- and $\bar{b}$-jet pair in the $H\to b\bar{b}$ (\textit{top row}) and $H\to gg$ (\textit{bottom row}) decay category using the flavour-$\kT$ (\textit{left column}) and flavour dressing (\textit{right column}) algorithm.}
    \label{fig:mbb}
\end{figure*}

%% General discussion of the plots.
Theoretical predictions for the four flavour-sensitive observables defined above are shown in \cref{fig:Ef1,fig:Ef2,fig:cosphi,fig:mbb} for two different jet-algorithm resolution parameter, $y_\mathrm{cut} = 0.1$ (solid lines) and $y_\mathrm{cut} = 0.01$ (dashed lines). In all figures, predictions in the $H\to b\bar b$ category are shown in the top row in red, while predictions for the $H\to gg$ category are shown in the bottom row in blue. Similarly, we show results using the flavour-$\kT$ algorithm in the left-hand column and results using the flavour-dressing algorithm in the right-hand column. Each plot consists of two panels, with the upper panel showing the NLO distributions and the lower panels showing the differential $K$-factor $\mathrm{NLO}/\mathrm{LO}$ for the two values of the jet-algorithm resolution parameter $y_\mathrm{cut}$. 

Scale variations are included by a lighter shaded band around the central predictions.

%% Discussion of general features and the infrared behaviour.
Generally, we observe larger rates in the $H\to b\bar{b}$ decay, owing to the fact that events in this category always contain at least one $b$-$\bar{b}$ pair, whereas most events in $H\to gg$ decays contain only unflavoured partons. 
We find larger NLO $K$-factors in the $H\to gg$ decay category, owing to the fact that our calculation employs the HEFT coupling between the Higgs and gluons, making it susceptible to large corrections. This is also reflected in the larger uncertainty band in predictions in the $H\to gg$ decay category.
In both Higgs decay categories, the magnitude of the NLO corrections is comparable between the flavour-$\kT$ and flavour-dressing algorithms, with somewhat larger corrections visible in the latter.
It can be seen that predictions calculated with larger values of $y_\mathrm{cut}$ generally also receive larger NLO corrections.
In all cases, lowering $y_\mathrm{cut}$ leads to wider distributions, owing to the larger kinematically allowed phase space.

Before moving on to discussing features specific to the individual observables, we wish to highlight a peculiar behaviour of the flavour-$\kT$ algorithm that results in some interesting phenomenology. For all observables shown here, distributions in the flavour-$\kT$ algorithm span a significantly larger value range than predictions employing the flavour-dressing algorithm. 
Although it might appear as if the flavour-$\kT$ algorithm allows to probe phase-space regions with infrared sensitive configurations, this is not the case. The reason for the large allowed value range can be found in the definition of the modified distance measure of the flavour-$\kT$ algorithm, \cref{eq:distmeas_flavour}, which takes the maximum whenever the softer of the two partons is flavoured.
In three-parton configurations $b\bar{b}g$ which contain a soft $b$-quark (or $\bar b$-quark), $E_b \ll E_g, E_{\bar b}$, cf.~\cref{fig:Hbbg}, all distances involving this quark involve the maximum of the energies, $y_{bj} \propto \max(E_b,E_j)$. As no singularity is associated with a single (anti-)quark becoming soft, this definition allows for arbitrarily small (anti-)quark energies, in principle below the order of the cut-off $y_\mathrm{cut}$, while still retaining $\min_{i,j}(y_{ij}) > y_\mathrm{cut}$. In other words, the flavour-$\kT$ algorithm counts three-particle configuration with a single arbitrarily soft quark as three-jet configurations, in contrast to the naïve expectation that a soft particle, regardless of its flavour, does not constitute a jet. We shall explore the consequence of this peculiar behaviour of the application of the flavour-$\kT$ algorithm in analysing thorougly the shape and normalisation of the distributions presented below.

%% Discussion of Ef1
\vspace{0.3cm}
\noindent\textbf{The energy of the leading flavoured jet}, $E_{b,1}$:\\
\Cref{fig:Ef1} shows the energy of the leading flavoured jet. 
In both decay categories and for both choices of the experimental resolution parameter $y_\mathrm{cut}$, we observe similar results in the distributions using the flavour-$\kT$ and flavour-dressing algorithm at lower energies. Towards the kinematical limit on the right-hand side of the plots, however, we find substantial differences between the two jet algorithms:
The flavour-dressing algorithm assigns a vanishing probability to configurations with $E_{b,1} \approx m_H/2$, whereas such configurations have a non-zero probability in the flavour-$\kT$ algorithm. This is related to the treatment of three-particle configurations containing a single soft quark in the flavour-$\kT$ algorithm.

For the $H \to b \bar b$ category we find the following $K$-factors for the two $y_{\mathrm{cut}}$ values: 
With a $y_{\mathrm{cut}}$ value of $0.1$  we find $K$-factors around $1.4-1.9$ for flavour-$\kT$ and  $1.3$ to $1.9$ for flavour-dressing. 

For $y_{\mathrm{cut}}=0.01$, we find $K$-factors ranging from $1.1$ ($1.2$) to $1.3$ ($1.3$) for flavour-$\kT$ (flavour-dressing).
In the $H \to gg$ category, we find higher NLO corrections, which are also reflected in the $K$-factors. For flavour-$\kT$ and a $y_{\mathrm{cut}}$ of $0.1$, we find $K$-factors from $1.5$ to $2.5$, where we start on the left side with a small increase from $2.4$ up to $2.5$ at an energy of $E_{b,1}/m_H=0.34$, followed by a decrease to $1.5$ for higher energies.
The $K$-factors are more constant for a $y_{\mathrm{cut}}$ value of $0.01$, varying around $1.4-1.7$.
Flavour-dressing with a $y_{\mathrm{cut}}$ of $0.1$ starts with $K$-factors of around $2.2$ and then rapidly increases. A similar behaviour is observed for a $y_{\mathrm{cut}}$ value of $0.01$, where we start at $1.7$ and then diverge again. Flavour-$\kT$ does not have these diverging NLO corrections, because of the previously discussed three-particle configurations containing a single soft quark. Analysing the situation regarding the size of the QCD corrections observed with both algorithms further, the following can be said: Already at LO, flavour-$\kT$ allows one quark to have large energies (and thus the other quark has almost no energy), whereas flavour-dressing does not allow for such configurations. Only at NLO, flavoured jets are allowed to have large energies in flavour dressing, and thus the $K$-factors receive large contributions there.

%% Discussion of Ef2
\vspace{0.3cm}
\noindent\textbf{The energy of the subleading flavoured jet}, $E_{b,2}$:\\
The distribution of the energy of the subleading flavoured jet, shown in \cref{fig:Ef2} can be explained by a similar reasoning as before for the leading flavoured jet distribution.
Due to the peculiar behaviour for pathological $b\bar{b}g$ configurations, the flavour-$\kT$ algorithm has a non-vanishing probability to find a flavoured jet with (almost) zero energy. As such however, the situation for the subleading flavoured jet is inverse to the situation of the leading flavoured jet. Indeed the corresponding subleading jet distribution now extends all the way to the maximum of $E_{b,1} \to m_H/2$, while $E_{b,2} \to 0$.
For this subleading jet distribution, the gluonic decay category is subject to rather large NLO corrections at higher energies. Understanding that flavoured quarks in the $H\to gg$ decay are exclusively generated by secondary gluon decays, it becomes clear that the high-energy tail of the subleading-energy distribution receives large corrections from real corrections in which both gluons decay to a $b$-$\bar b$ pair.
The $K$-factors for the fermionic decay category have a small increase for very low energies in the flavour-$\kT$ algorithm and a decrease for high-energies for both algorithms. The ranges are from $1.3$ ($1.2$) at  to $2.0$ ($1.9$) for flavour-$\kT$ with $y_{\mathrm{cut}}=0.1$ ($y_{\mathrm{cut}}=0.01)$. For flavour-dressing the ranges are  from $1.3$ ($1.3$) to $1.5$ ($1.4)$ with $y_{\mathrm{cut}}=0.1$ ($y_{\mathrm{cut}}=0.01$). The increase is for all algorithms and $y_{\mathrm{cut}}$ values towards lower energies.
The gluonic channel has diverging $K$-factors for high energies, as alluded to above. Otherwise, the $K$-factors range for a $y_{\mathrm{cut}}$ of $0.1$ from $2.1$ to $3.0$ for flavour-$\kT$ and $2.6$ to $3.0$ for flavour-dressing, where for both the $K$-factor of $3.0$ is obtained at an energy of $E_{b,2}/m_H=0.3$. For a $y_{\mathrm{cut}}$ of $0.01$ the ranges are from $1.7$ to $1.8$ for flavour-$\kT$ and around $2.0-2.8$ for flavour dressing, where again the upper value is obtained at an energy of $E_{b,2}/m_H=0.3$. Afterwards the $K$-factors diverge, as alluded to above.

\vspace{0.3cm}
\noindent\textbf{The angular separation of the leading $b$- and $\bar b$-jet}, $\cos\theta_{b\bar b}$:\\
\Cref{fig:cosphi} shows the angular separation of the leading $b$- and leading $\bar b$-jet. We see that the peak of the distributions is located at large angles ($\cos\phi \to -1$) in the $H \to b \bar b$ category, whereas it is located at small angles ($\cos \phi \to 1$) in $H\to gg$ decays. 
The reason is that in Yukawa-induced $H\to b\bar{b}$, the quark-antiquark pair directly stems from the Higgs decay and is thus expected to favour a large angular separation, whereas quark-antiquark pairs stem from gluon decays in the $H\to gg$ category and, as such, are divergent in the collinear limit.
In $H\to gg$ decays, we see another peak towards $\cos\phi = -1$, which originates from decays $H \to gg \to b \bar b b \bar b$, in which the leading $b$ and leading $\bar b$ originate from different gluons. Since this only happens at NLO, these obtain large NLO-corrections. The same is not visible in $H\to b\bar b$ decays, as the leading $b$-$\bar b$ pair almost exclusively stems from the primary Higgs-decay vertex.
The peculiar behaviour of the flavour-$\kT$ algorithm for the three-parton configurations with a single soft $b$-quark again allows the distributions to span further into the infrared region, as is visible by the numerically larger upper limit in the distributions obtained with flavour-$\kT$.
Except for the peak towards $\cos\phi=-1$, the $K$-factors for the $H \to b \bar b$ channel are ranging from $1.3$ ($0.8)$ to $1.7$ ($1.3$) for flavour-$\kT$ and a $y_{\mathrm{cut}}$  of $0.1$ ($0.01$), while the range for flavour-dressing with a $y_{\mathrm{cut}}$ of $0.1$ is from $1.2$ to $1.6$, where the decrease is towards higher $\cos\phi$. The distribution is ended by a sudden increase in the NLO corrections. Flavour-dressing with a $y_{\mathrm{cut}}$ of $0.01$ starts around a $K$-factor of $1.3$ at low energies and then slowly decreases to a $K$-factor of $1.0$ for higher energies.
For $H\to gg$ we find the already discussed explosion of NLO correction towards $\cos\phi=-1$ and have otherwise $K$-factors in a range from $2.1$ ($1.5$) to $2.2$ ($1.9$) for a $y_{\mathrm{cut}}$ value of $0.1$ ($0.01$), where for $y_{\mathrm{cut}}=0.01$ the minimal $K$-factor of $1.5$ is obtained around $\cos\phi = 0$ and possesses a steady increase of the NLO corrections in both directions.
Flavour-dressing again features huge NLO corrections towards $\cos\phi=-1$. The descent for a $y_{\mathrm{cut}}$ of $0.1$ leads at $\cos\phi=-0.5$ to a $K$-factor of $3.0$, and yields at $\cos\phi=0$ a $K$-factor of $2.4$. It then falls further to around $2.2$ followed by a sharp increase again.
For a $y_{\mathrm{cut}}$ of $0.01$, the $K$-factors have a flatter distribution, where after the explosion on the left-hand side, it reaches at $\cos\phi=-0.5$ a $K$-factor of $2.3$ and slowly decreases afterwards down to a $K$-factor of $1.7$ followed again by a sharp spike.

\vspace{0.3cm}
%% Discussion of mbb
\noindent\textbf{The invariant mass of the leading $b$- and $\bar b$-jet}, $m_{b\bar b}$:\\
\Cref{fig:mbb} shows the invariant mass of the leading $b$- and leading $\bar b$-jet, $m_{b\bar b}$. 
While Yukawa-induced decays favour quarks with a high invariant mass, $H\to gg$ decays favour quark-antiquark pairs with a small invariant mass.
Again this can be understood by noting that the quarks in $H\to b\bar b$ decays originate directly from the primary Higgs-decay vertex, whereas they stem from secondary gluon splittings in $H\to gg$ decays and will thus have a smaller invariant mass $m_{b\bar b} \sim (1-\cos\theta_{b\bar{b}})$, due to the form of the $g\to q\bar q$ splitting amplitude.
The difference becomes more pronounced for lower values of $y_\mathrm{\mathrm{cut}}$, where the distribution shifts to larger angles in Yukawa-induced decays and towards smaller angles in gluonic decays.
The main difference between the jet algorithms is that the flavour-$\kT$ allows for smaller values of $m_{b \bar b}$, which again is explained by the maximum in the distance measure in equation \eqref{eq:distmeas_flavour}.
The $K$-factors for a $y_{\mathrm{cut}}$ of $0.1$ in the fermionic decay category start by a peak at low invariant mass, where the peak for the flavour-dressing goes below a $K$-factor of $1$. Both algorithms then do not have too much variation in the $K$-factors. For flavour-$\kT$ the values range from $1.5$, at higher invariant mass, to $1.6$ at lower invariant mass. In flavour-dressing the increase is in the opposite direction with values ranging from $1.3$, obtained around $m_{b \bar b}=0.4$, to $1.5$ for higher invariant mass. Both algorithms then end the distributions with another peak.
Lowering the $y_{\mathrm{cut}}$ value to $0.01$ again flattens the distribution a bit. Flavour-$\kT$ starts for low invariant mass at a $K$-factor of $1.1$ and then increases to a maximum of $1.4$. Thereafter it slowly decreases to a $K$-factor of $1.3$ for higher invariant mass.
The flavour-dressing algorithm has a small $K$-factor of $0.2$ for very low invariant mass, but then quickly increases again until reaching a $K$-factor of $1.2$ around $m_{b \bar b}/m_H=0.3$ followed by further increase to $1.3$ for higher invariant mass.

The $H\to gg$ decay channel has big variations in the $K$-factors and need a careful analysis. Starting with flavour-$\kT$ and $y_{\mathrm{cut}}=0.1$, we observe that the NLO corrections diverge towards negative infinity for small invariant mass, leaving thereby the region where the fixed order prediction can be trusted.
Increasing the invariant mass also increases the NLO corrections, and the $K$-factor crosses a value of $1.0$ around $m_{b \bar b}/m_H = 0.03$ and further rapidly increases to $1.8$ around $m_{b \bar b}/m_H = 0.1$, $2.4$  at $m_{b \bar b}/m_H = 0.3$, $2.7$ at $m_{b \bar b}/m_H = 0.5$ and further to $4.0$ at $m_{b \bar b}/m_H = 0.7$, where it then starts to diverge towards positive infinity. 

The distribution is similar for flavour-$\kT$ with a $y_{\mathrm{cut}}$ value of $0.01$ with the main difference, that after reaching a peak in the $K$-factor of $2.3$ around $m_{b \bar b}/m_H=0.16$, it starts to decrease again to a $K$-factor of $1.7$, before the NLO-corrections get larger again for higher invariant mass.
For flavour-dressing the distributions are similar to flavour-$\kT$, where again for $y_{\mathrm{cut}}=0.1$, we start with very small NLO-corrections at low invariant mass followed by a quick increase and a divergence towards positive infinity. Similarly for $y_{\mathrm{cut}}=0.01$, the NLO-corrections are low at small invariant mass. After a rapid increase to a $K$-factor of $1.8$ at $m_{b \bar b}/m_H=0.2$, the NLO-corrections increases to $2.8$ at $m_{b \bar b}/m_H=0.6$ and then start to diverge again.

\section{Summary and outlook}\label{sec:conclusions}
In this paper, for the first time, we have computed flavour-sensitive observables related to hadronic Higgs decays including both Higgs decay categories, i.e stemming from the underlying processes $H\to b\bar b$ and $H\to gg $ including higher order corrections up to $\mathcal{O}(\alphaS^2)$ in perturbative QCD. 
The calculation was carried out in an effective theory in which the Higgs couples directly to gluons, while massless $b$-quarks retain a non-vanishing Yukawa coupling.
Specifically, we considered the following observables : the energy of the leading and subleading flavoured jet, the angular separation and the invariant mass of the leading $b$-$\bar{b}$ pair at relative $\mathcal{O}(\alphaS^2)$ in QCD.
Using the antenna-subtraction framework, the computations were performed with the parton-level event generator \eerad, extended to account for hadronic Higgs decays. A new flavour layer was 
implemented in \eerad to allow for the computation of observables with identified flavoured jets.
Flavoured jets were defined using both the flavour-$\kT$ and flavour-dressing algorithm. For both algorithms, infrared flavour safety was explicitly verified in \cref{subsec:flavour_safety}.
For each observable, predictions obtained in both Higgs-decay categories, using both flavoured jet algorithms and for two values of the experimental resolution parameter $y_\mathrm{cut}$, i.e., $y_{\mathrm{cut}}=0.1$ and $y_{\mathrm{cut}}=0.01$ were compared.
In all cases, lowering the parameter $y_\mathrm{cut}$ leads to wider distributions, owing to the larger kinematically allowed phase space.

Comparing the two Higgs decay modes, we observe larger rates in the $H\to b\bar{b}$ decay category. This is related to the fact that events in this category always contain at least one $b$-$\bar{b}$ pair, whereas most events in $H\to gg$ decays contain only unflavoured partons. 
We find larger NLO $K$-factors in the $H\to gg$ decay category, owing to the fact that our calculation employs the HEFT coupling between the Higgs and the gluons.
Comparing the results obtained using the two flavoured jet algorithms, it was highlighted that the use of the flavour-$\kT$ algorithm introduces counter-intuitive phenomenological implications, owing to its treatment of pathological three-parton configurations containing a single soft flavoured quark. While states with a single soft quark do not correspond to infrared singularities in physical matrix elements, it has the effect that these configurations are identified as three-jet like despite containing an, in principle, arbitrarily soft quark.

Away from phase-space regions dominated by configurations with a single soft flavoured quark, we find qualitatively good agreement between the flavour-$\kT$ and the flavour-dressing algorithm, with generally slightly larger NLO corrections in the latter.
Using the flavour dressing algorithm in particular, the shape and normalisation of the individual distributions in both decay categories have characteristic fixed-order behaviour over the whole kinematical range. In particular the behaviour at the kinematical edges, i.e., the drop or peak in the cross section, can be understood considering only hard final states in both Higgs decay categories. It is worth mentioning though, that the drop or the peak of the distributions seen at the kinematical edges are systematically exchanged in one or the other Higgs decay categories.  
This analysis demonstrates in particular the practical applicability of the flavour-dressing algorithm to compute flavour-sensitive observables in hadronic Higgs decays including both decay modes.

Our study marks the first step towards a more complete treatment of flavour-induced effects in hadronic Higgs decays. Obtaining a solid theoretical understanding of these effects will be vital for Higgs precision studies at future lepton colliders. 
Among these efforts, two avenues for future work are particularly worth mentioning: 
the study of flavour-tagged event shapes 
and the analysis of the phenomenological impact gained by the inclusion of NNLO-type corrections to flavour-sensitive observables in hadronic Higgs decays.  
We anticipate to return to both avenues in the future.

\begin{acknowledgements}
We would like to thank Giovanni Stagnitto for useful discussions and a careful reading of the manu-script.
AG and CTP acknowledge support by the Swiss National Science Foundation (SNF) under contract 200021-197130 
and by the Swiss National Supercomputing Centre (CSCS) under project ID ETH5f.
BCA is supported by the Deutsche Forschungsgemeinschaft (DFG, German Research Foundation) under grant 396021762 - TRR257.
Parts of the computations were carried out on the PLEIADES cluster at the University of Wuppertal, supported by the Deutsche Forschungsgemeinschaft (DFG, grant No. INST 218/78-1 FUGG) and the Bundesministerium für Bildung und Forschung (BMBF).
\end{acknowledgements}

\bibliographystyle{spphys}       % APS-like style for physics
\bibliography{bibliography}

\end{document}